\documentclass[12pt] {article}
\usepackage{psfrag}
\usepackage{graphicx}
\usepackage{latexsym,amsfonts}
\usepackage{amssymb}
\begin{document}
\setcounter{page}{1}
\def\theequation{\arabic{section}.\arabic{equation}}
\def\theequation{\thesection.\arabic{equation}}
\setcounter{section}{0}

\title{Phenomenological model of the Kaonic Nuclear Cluster $K^-pnn$
  in the ground state}

\author{A. N. Ivanov$^{1,3}$\,\thanks{E--mail:
    ivanov@kph.tuwien.ac.at, Tel.: +43--1--58801--14261, Fax:
    +43--1--58801--14299}~\thanks{Permanent Address: State Polytechnic
    University, Department of Nuclear Physics, 195251 St. Petersburg,
    Russian Federation}\,, P.  Kienle$^{2,3}$\thanks{E--mail:
    Paul.Kienle@ph.tum.de}\,, J.
  Marton$^3$\thanks{E--mail:johann.marton@oeaw.ac.at}\,, E.
  Widmann$^3$\,\thanks{E--mail: eberhard.widmann@oeaw.ac.at,
http://www.oeaw.ac.at/smi}}

\date{\today}

\maketitle

\begin{center} {\it $^1$Atominstitut der \"Osterreichischen
    Universit\"aten, Technische Universit\"at Wien, Wiedner
    Hauptstrasse 8-10, A-1040 Wien, \"Osterreich, \\
    $^2$Physik Department, Technische Universit\"at M\"unchen,
    D--85748 Garching, Germany,\\ $^3$Stefan Meyer Institut f\"ur
    subatomare Physik, \"Osterreichische Akademie der Wissenschaften,
    Boltzmanngasse 3, A-1090, Wien, \"Osterreich}
\end{center}

\begin{center}
\begin{abstract}
  A phenomenological model, proposed for the Kaonic Nuclear Cluster
  (KNC) $K^- pp$ (or ${^2_{\bar{K}}}{\rm H}$), is applied to the
  theoretical analysis of the KNC $K^-pnn$ (or $S^0$). The theoretical
  values of the binding energy and the width are equal to
  $\epsilon_{S^0} = -\,197\,{\rm MeV}$ and $\Gamma_{S^0} = 16\,{\rm
    MeV}$. They agree well with the experimental data
  $\epsilon^{\exp}_{S^0} = -\,194.0^{+\,1.5}_{-\,4.4}\,{\rm MeV}$ and
  $\Gamma^{\exp}_{S^0} < 21\,{\rm MeV}$ (PLB {\bf 597}, 263 (2004))
  and the theoretical values $\epsilon_{S^0} = -\,191\,{\rm MeV}$ and
  $\Gamma_{S^0} = 13\,{\rm MeV}$, obtained by Akaishi and Yamazaki
  within the potential model approach.
\end{abstract}

PACS: 11.10.Ef, 13.75.Jz, 14.20.Jn, 25.80.Nv

\end{center}

\newpage

\section{Introduction}
\setcounter{equation}{0}

Recently \cite{TS04} Suzuki {\it et al.} have announced the
experimental discovery of the bound state $K^-pnn$, named $S^0(3115)$,
with quantum numbers $I(J^P) = 1(\frac{3}{2}^+)$, the binding energy
$\epsilon^{\exp}_{S^0} = -\,194.0^{+\,1.5}_{-\,4.4}\,{\rm MeV}$ and
the width constrained by $\Gamma^{\exp}_{S^0} < 21\,{\rm MeV}$.  The
possible existence of such a Kaonic Nuclear Cluster (KNC) as well as
of the KNC $K^-pp$ (or ${^2_{\bar{K}}}{\rm H}$), which has been
observed by the FINUDA Collaboration \cite{EDB1}, has been pointed out
by Akaishi and Yamazaki within the potential model approach
\cite{TDB1}--\cite{EDB2}.  According to \cite{TS04}, the main decay
channels of the KNC $S^0(3115)$ are $S^0 \to \Sigma^- pn$, $S^0 \to
\Sigma^- d$ and $S^0 \to \Sigma^0 nn$, where $d$ is the deuteron. The
decay channel $S^0 \to \Lambda^0 nn$ is suppressed.

In this paper we extend the phenomenological quantum field theoretic
model for the KNC ${^2_{\bar{K}}}{\rm H}$, proposed in \cite{SMI1}, to
the description of the KNC $S^0(3115)$. The binding energy and the
width of the KNC $S^0(3115)$ we define as \cite{SMI1}
\begin{eqnarray}\label{label1.1}
  -\,\epsilon_{S^0} + i\,\frac{\Gamma_{S^0}}{2} = 
  \int d\tau\,\Phi^*_{S^0}\,M(K^-pnn \to 
K^-pnn)\,\Phi_{S^0},
\end{eqnarray}
where $d\tau$ is an element of the phase space of the $K^-pnn$ system,
$\Phi_{S^0}$ is the wave function of the KNC $S^0(3115)$ and $M(K^-pnn
\to K^-pnn)$ is the amplitude of $K^-pnn$ scattering.

In our model for the calculation of the scattering amplitude of the
constituents of the KNC we use the chiral Lagrangian with a
non--linear realization of chiral $SU(3)\times SU(3)$ symmetry and
derivative meson--baryon couplings \cite{BWL68}, accepted for the
analysis of strong low--energy interactions in ChPT \cite{JG83} (see
also \cite{HL00,EK01}). We calculate these amplitudes in the
tree--approximation to leading order in the large $N_C$ and chiral
expansion, where $N_C$ is the number of quark colour degrees of
freedom \cite{EW79}. According to Witten \cite{EW79}, the large $N_C$
expansion is equivalent to the heavy--baryon approximation, applied in
ChPT \cite{JG83}--\cite{EK01} to the analysis of low--energy
meson--baryon interactions (see also \cite{TE05,AI99}).

In our model the main part of correlations between constituents of the
KNC are described by the wave function of the KNC. 

For the construction of the wave function $\Phi_{S^0}$ one has to
specify the structure of the KNC $S^0(3115)$.  The simplest structure
of the bound $K^-pnn$ system with quantum numbers $I(J^P) =
1(\frac{3}{2}^+)$ and the decay mode $S^0 \to \Lambda^0 nn$ suppressed
is $(K^-p)_{I = 0}(nn)^{\ell = 1}_{I = 1}$, where the $K^-p$ pair is
in the state with isospin $I = 0$, i.e.  $(K^-p)_{I = 0}$, and the
$nn$ pair is in the P--wave and spin--triplet state with isospin $I =
1$, i.e. $(nn)^{\ell = 1}_{I = 1}$.

The paper is organised as follows. In Section 2 we calculate the
binding energy of the KNC $S^0(3115)$. We get $\epsilon_{S^0} = -
197\,{\rm MeV}$. According to \cite{SMI1}, the dominant contribution
comes from the Weinberg--Tomozawa term for $K^-pnn$ scattering. In
Section 3 we calculate the partial widths of the decay modes $S^0 \to
\Sigma^- pn $, $S^0\to \Sigma^- d$, $S^0 \to \Sigma^0 nn$ and $S^0 \to
\Lambda^0 nn$ and the total width of the KNC $S^0(3115)$.  We get
$\Gamma_{S^0} = 16\,{\rm MeV}$. We show that the dominant decay mode
is $S^0 \to \Sigma^- pn $ with the partial width $\Gamma(S^0 \to
\Sigma^- pn) = 15\,{\rm MeV}$. We calculate the invariant--mass
spectrum of the $S^0 \to \Sigma^- p n$ decay. In the Conclusion we
discuss the obtained results.  In the Appendix we define the wave
function $\Phi_{S^0}$ of the KNC $S^0(3115)$.

\section{Binding energy of the KNC $S^0(3115)$}
\setcounter{equation}{0}

According to \cite{SMI1}, the binding energy of the KNC $S^0(3115)$
is given by
\begin{eqnarray}\label{label2.1}
  -\,\epsilon_{S^0} =
  \frac{5}{4}\,\frac{1}{F^2_{\pi}}\,
\Big(\frac{\mu^3_1\omega^2_1\Omega_1}{\pi^3}\Big)^{1/2} 
= 197\,{\rm MeV},
\end{eqnarray}
where $F_{\pi} = 92.4\,{\rm MeV}$ is the PCAC constant of pseudoscalar
mesons. Other parameters $\mu_1$, $\omega_1$ and $\Omega_1$ are
related to the wave function $\Phi_{S^0}$ of the KNC $S^0(3115)$ and
defined in the Appendix.  The theoretical value agrees well with the
experimental one $\epsilon^{\exp}_{S^0} =
-\,194.0^{+\,1.5}_{-\,4.4}\,{\rm MeV}$.  The binding energy is defined
by the contribution of the Weinberg--Tomozawa term. This term gives
the dominant contribution to the real part of the amplitude of elastic
$K^-pnn$ scattering to leading order in the large $N_C$ and chiral
expansion.

The correction of next--to--leading order in the large $N_C$ and
chiral expansion to the Weinberg--Tomozawa term is defined by the
diagrams in Fig.\,1.  The calculation of diagrams in Fig.\,1 is
carried out using the chiral Lagrangian with a non--linear realization
of chiral $SU(3)\times SU(3)$ symmetry and derivative meson--baryon
couplings \cite{SMI1}.  For $N_C = 3$ such a correction makes up a few
percent of the Weinberg--Tomozawa term.

\begin{figure}
  \centering \psfrag{K-}{$K^-$}
 \psfrag{p}{$p$}
\psfrag{n}{$n$}\psfrag{p0}{$\pi^0$} 
\psfrag{a}{$+$}
\psfrag{b}{$+ ~\ldots$}
\includegraphics[height= 0.14\textheight]{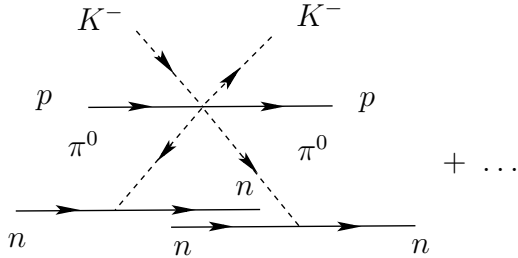}
\caption{Feynman diagrams giving in the tree--approximation
  contributions to the amplitude of the reaction $K^-pnn \to K^-pnn$
  to next--to--leading order in the large $N_C$ and chiral expansion
  with respect to the Weinberg--Tomozawa term.}
\end{figure}
\section{Width of the KNC $S^0(3115)$}
\setcounter{equation}{0}

In this section we calculate the partial widths of the decay modes
$S^0 \to \Sigma^- pn$, $S^0 \to \Sigma^- d $, $S^0 \to \Sigma^0 nn$
and $S^0 \to \Lambda^0 nn$ and the total width $\Gamma_{S^0}$ of the
KNC $S^0(3115)$. For the total width we get the value $\Gamma_{S^0} =
16\,{\rm MeV}$, agreeing well with the experimental data \cite{TS04}
$\Gamma_{S^0} < 21\,{\rm MeV}$ and the potential model prediction
$\Gamma_{S^0} = 13\,{\rm MeV}$ \cite{TDB2} (see also \cite{EDB2}).

\subsection{The $S^0 \to \Sigma ^- p n$ decay
  mode}

The Feynman diagrams of the reaction $K^-pnn \to \Sigma^- pn $ are
depicted in Fig.\,2. In the tree--approximation these diagrams give
the main contribution to the amplitude of the $S^0 \to \Sigma^-pn $
decay to leading order in the large $N_C$ and chiral expansion. The
calculation of these diagrams is carried out using the chiral
Lagrangian with chiral $SU(3)\times SU(3)$ symmetry and derivative
meson--baryon couplings (see Appendix B of Ref.\cite{SMI1}). The
partial width of the $S^0\to \Sigma^-pn$ decay is equal to
\begin{eqnarray}\label{label3.1}
 \hspace{-0.3in}\Gamma(S^0 \to \Sigma^- p n) &=& 
  \Big(\frac{\mu^3_1\omega^2_1\Omega_1}{\pi^3}\Big)^{\!1/2}
  \Big(\frac{\mu_2\omega_2}{\pi}\Big)^{\!3/2}
  \Big(\frac{\mu_3\omega_3}{\pi}\Big)^{\!5/2}\frac{9}{32\pi^2}
\,\frac{g^2_{\pi NN}}{m_K m^4_{\pi}F^8_{\pi}}\,\sqrt{\frac{2 \mu^3}{ m_N}}\, 
\nonumber\\
\hspace{-0.3in}&& \times\,\varepsilon^2
\,f(\varepsilon) = 15\,{\rm MeV},
\end{eqnarray}
where $g_{\pi NN} = g_A m_N/F_{\pi} = 13.3$ is the $\pi NN$ coupling
constant \cite{PSI2,TE04}, $\varepsilon = m_{S^0} - 2m_N - m_{\Sigma}
= 44\,{\rm MeV}$ is the energy excess calculated for $m_N = 940$ and
$m_{\Sigma} =1193\,{\rm MeV}$, $\mu = 2m_N m_{\Sigma}/(2 m_N +
m_{\Sigma}) = 730\,{\rm MeV}$ and $f(\varepsilon) = 0.78$ is the
contribution of the phase volume of the $\Sigma^-pn$ state. The
parameters $\mu_i$, $\omega_i$ for (i=1,2,3) and $\Omega_1$ of the
wave function $\Phi_{S^0}$ of the KNC $S^0(3115)$ are defined in the
Appendix.
\begin{figure}
 \psfrag{K-}{$K^-$} 
\psfrag{K0}{$\bar{K}^0$} 
\psfrag{S0}{$\Sigma^0$}
\psfrag{S-}{$\Sigma^-$}
\psfrag{p}{$p$}
\psfrag{n}{$n$}
\psfrag{p-}{$\pi^-$}
\psfrag{p+}{$\pi^+$}
\psfrag{p0}{$\pi^0$}
 \psfrag{b}{$+ ~\ldots$}
\psfrag{a}{$+$} 
\psfrag{A}{$(A)$}
\psfrag{B}{$(B)$}
\psfrag{C}{$(C)$}
\psfrag{D}{$(D)$}
\includegraphics[height= 0.120\textheight]{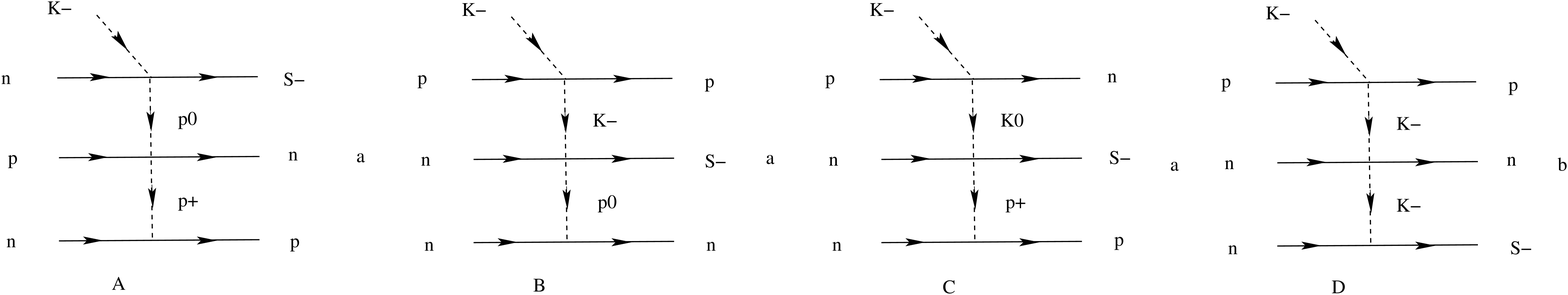}
\caption{Feynman diagrams of the amplitude of the reaction $K^-pnn \to
  \Sigma^- pn$ giving in the tree--approximation the main contribution
  to leading order in the large $N_C$ and chiral expansion.}
\end{figure}

\subsection{Invariant--mass spectrum of the $S^0 \to \Sigma^- p
  n$ decay }

Following \cite{PK06} we calculate the invariant--mass spectrum of the
$S^0 \to \Sigma^- p n$ decay.  For this aim we use the following
variables \cite{PK06}
\begin{eqnarray}\label{label3.2}
 X = m^2_{np} = (E_n + E_p)^2 - (\vec{k}_n + \vec{k}_p)^2\;,\;
Y =  m^2_{\Sigma p} = (E_{\Sigma^-} + E_p)^2 - (\vec{k}_{\Sigma} + \vec{k}_p)^2,
\end{eqnarray}
which have the meaning of the invariant--squared masses of the $np$
and $\Sigma^-p$ pairs, respectively. At the rest frame of the KNC
$S^0(3115)$ in terms of the variables $(X,Y)$ the kinetic energies of
the neutron and the proton are given by
\begin{eqnarray}\label{label3.3}
T_n = \frac{(m_{S^0} - m_N)^2 - Y}{2 m_{S^0}}\;,\;
T_p  =  \varepsilon - \frac{(m_{S^0} - m_{\Sigma})^2 + (m_{S^0} - m_N)^2 - X - Y}{2 m_{S^0}}.
\end{eqnarray}
The invariant--mass spectrum or the Dalitz density distribution of the
$S^0 \to \Sigma^- p n$ decay is defined by
\begin{eqnarray}\label{label3.4}
  m^3_{S^0}\,\frac{d^2\Gamma(S^0 \to \Sigma^- p n)}{dX dY} &=& 
\Big(\frac{\mu^3_1\omega^2_1\Omega_1}{\pi^3}\Big)^{\!1/2}
  \Big(\frac{\mu_2\omega_2}{\pi}\Big)^{\!3/2}
  \Big(\frac{\mu_3\omega_3}{\pi}\Big)^{\!5/2}\,\frac{9}{128\pi^2}\,
\frac{g^2_{\pi NN} m_{\Sigma} m_{S^0}}{m_K m^4_{\pi} F^8_{\pi}}\nonumber\\
&&\times\, [A(X,Y) + B(X,Y) + C(X,Y) + D(X,Y)]^2 = \nonumber\\
&=& 10\,[A(X,Y) + B(X,Y) + C(X,Y) + D(X,Y)]^2.
\end{eqnarray}
In the $(X,Y)$--plane the Dalitz domain is bounded by the curve with
$X_{\rm min} = 4m^2_N$, $X_{\rm max} = (m_{S^0} - m_{\Sigma})^2$,
$Y_{\rm min} = (m_{\Sigma} + m_N)^2$ and $Y_{\rm max} = (m_{S^0} -
m_N)^2$ \cite{PK06}.
\begin{figure}
 \centering
\psfrag{X}{$X = m^2_{np}({\rm GeV}^2)$} 
\psfrag{Y}{$Y = m^2_{\Sigma p}({\rm GeV}^2)$} 
\includegraphics[height= 0.27\textheight]{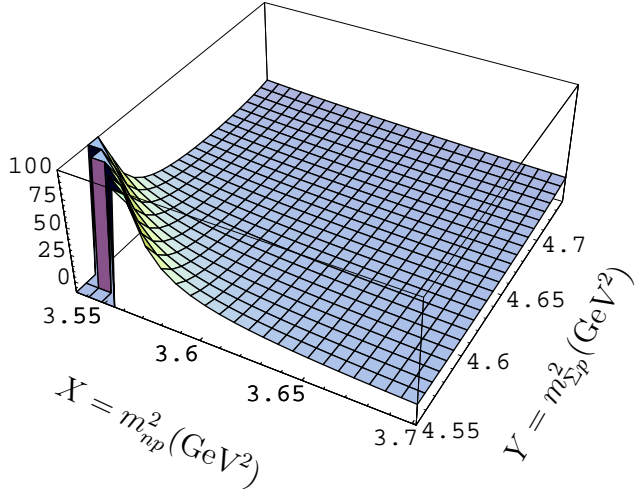}
\caption{The invariant--mass spectrum of the $S^0 \to \Sigma^- p
  n$ decay in the variables $X = m^2_{np}$ and $Y = m^2_{\Sigma p}$
  \cite{PK06}.}
\end{figure}
The invariant--mass spectrum of the $S^0 \to \Sigma^- p n$ is
represented in Fig.\,3.

\subsection{The $S^0 \to \Sigma^- d$ decay
  mode}
The partial width of the $S^0 \to \Sigma^- d$ decay is equal to
\cite{IV4}
\begin{eqnarray}\label{label3.5}
  \hspace{-0.3in}&&\Gamma(S^0 \to \Sigma^-d) =
  \Big(\frac{\mu^3_1\omega^2_1\Omega_1}{\pi^3}\Big)^{\!1/2}
  \Big(\frac{\mu_2\omega_2}{\pi}\Big)^{\!3/2}
  \Big(\frac{\mu_3\omega_3}{\pi}\Big)^{\!5/2}\,
  \frac{9\,g^2_{\pi NN} m_{\Sigma} Q_{\Sigma d}}{4 m_K m_N m_{S^0} m^4_{\pi}F^8_{\pi}}\nonumber\\
  \hspace{-0.3in}&&\times\Big|\int \frac{d^3 k}{(2\pi)^3}
\frac{\Phi_d(k)}{\displaystyle 1 -
    \frac{1}{2}r^t_{np}a^t_{np}k^2 - i\,a^t_{np} k}
  (A + B + C + D)\Big|^2 =  1\,{\rm MeV},
\end{eqnarray}
where the amplitudes $A$, $B$, $C$ and $D$ are defined by the
corresponding Feynman diagrams in Fig.\,2, $Q_{\Sigma d} = 266\,{\rm
  MeV}$ is a relative momentum of the $\Sigma^- d$ pair, $\Phi_d(k) =
\sqrt{8\pi \gamma}/(\gamma^2 + k^2)$ is the wave function of the
deuteron in the ground state in the momentum representation and
$\gamma = 1/R_d = 46\,{\rm MeV}$ \cite{MN79}. The amplitude of the
$np$ scattering is defined according to \cite{FSI} (see also
\cite{IV4}) and taken in the effective range approximation
\begin{eqnarray}\label{label3.6}
 k\cot\delta_{np}(k) = -\,\frac{1}{a^t_{np}} + \frac{1}{2}\,r^t_{np}k,
\end{eqnarray}
where $\delta_{np}(k)$ is the phase shift of the $np$ scattering in
the ${^3}{\rm S}_1$ state, $a^t_{np} = (5.424 \pm 0.004)\,{\rm fm}$
and $r^t_{np} = (1.759 \pm 0.005)\,{\rm fm}$ are the S--wave
scattering length and effective range of $np$ scattering in the
${^3}{\rm S}_1$ state \cite{MN79}. 

The main contribution to the integral comes from the relative momenta
$k \sim 100\,{\rm MeV}$, and 94$\,\%$ of the value of the
integral are concentrated in the region $0\le k \le 200\,{\rm MeV}$.
For the relative momenta $k = 100\,{\rm MeV}$ and $ k = 200\,{\rm
  MeV}$ the phase shift of the $np$ scattering, defined by the
effective range approximation (\ref{label3.6}), is equal to
$\delta_{np} = 85.3^0$ and $\delta_{np} = 54.6^0$, respectively.
These values agree well with the SAID analysis of the experimental
data on the $np$ scattering in the ${^3}{\rm S}_1$ state \cite{SAID}:
$\delta_{np} = 85.84^0$ and $\delta_{np} = 48.84^0$, respectively.

\subsection{The $S^0 \to \Sigma^0 nn$ and $S^0 \to \Lambda^0 nn$ decay
  modes }
\begin{figure}
 \centering
\psfrag{K-}{$K^-$} 
\psfrag{K0}{$\bar{K}^0$} 
\psfrag{S0}{$\Sigma^0$}
\psfrag{S-}{$\Sigma^-$}
\psfrag{Y}{$\Lambda^0,\Sigma^0$}
\psfrag{p}{$p$}
\psfrag{n}{$n$}
\psfrag{p-}{$\pi^-$}
\psfrag{p+}{$\pi^+$}
\psfrag{p0}{$\pi^0$}
 \psfrag{b}{$+ ~\ldots$}
\psfrag{a}{$+$} 
\includegraphics[height= 0.115\textheight]{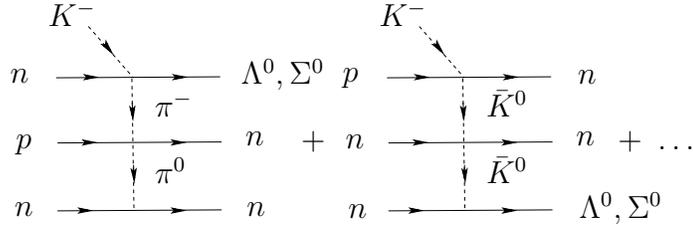}
\caption{Feynman diagrams of the amplitudes of the reactions $K^-pnn
  \to \Lambda^0 nn$ and $K^-pnn \to \Sigma^0 nn$ giving in the
  tree--approximation the main contribution to leading order in the
  large $N_C$ and chiral expansion.}
\end{figure}
The amplitudes of the $S^0\to \Sigma^0 nn$ and $S^0 \to \Lambda^0 nn$
decays are defined by the integrals of the amplitudes of the reactions
$K^-pnn \to \Sigma^0 nn$ and $K^-pnn \to \Lambda^0 nn$ with the wave
function $\Phi_{S^0}$ of the KNC $S^0(3115)$. Feynman diagrams of the
amplitudes of the reactions $K^-pnn \to \Sigma^0 nn$ and $K^-pnn \to
\Lambda^0 nn$ are depicted in Fig.4. In the tree--approximation these
diagrams give the main contribution to leading order in the large
$N_C$ and chiral expansion.  The contributions of the amplitudes of
the reactions $K^-pnn \to \Sigma^0 nn$ and $K^-pnn \to \Lambda^0 nn$,
calculated to leading order in the large $N_C$ and chiral expansion,
to the amplitudes of the $S^0 \to \Sigma^0 nn$ and $S^0\to \Lambda^0
nn$ decays vanish. This means that the decay modes $S^0 \to \Sigma^0
nn$ and $S^0\to \Lambda^0 nn$ are suppressed to leading order in the
large $N_C$ and chiral expansion.

\subsection{Width of the KNC $S^0(3115)$  }

The width of the KNC $S^0(3115)$ is defined by the sum of the partial
widths of all modes. Since the dominant modes are $S^0 \to \Sigma^- p
n$ and $S^0 \to \Sigma^- d$, the width of the $S^0(3115)$ is equal to
\begin{eqnarray}\label{label3.7}
  \Gamma_{S^0} = 16\,{\rm MeV}.
\end{eqnarray}
This agrees well with the experimental constraint $\Gamma^{\exp}_{S^0}
< 21\,{\rm MeV}$ and the theoretical value $\Gamma_{S^0} = 13\,{\rm
  MeV}$, predicted by Akaishi and Yamazaki within the potential model
approach \cite{TDB2} (see also \cite{EDB2}).

\section{Conclusion}

The phenomenological quantum field theoretic model of the KNC $K^-pp$
(or ${^2_{\bar{K}}}{\rm H}$), proposed in \cite{SMI1}, is extended to
the description of the KNC $S^0(3115)$ with quantum numbers $I(J^P) =
1(\frac{3}{2}^+)$ \cite{TS04}.  The wave function of the KNC
$S^0(3115)$ is taken in the oscillator form. The frequencies of
oscillations are related to the frequency $\omega$, describing
longitudinal oscillations of the $K^-$--meson with respect to the $pp$
pair in the KNC ${^2_{\bar{K}}}{\rm H}$ \cite{SMI1}. The value of the
frequency $\omega = 64\,{\rm MeV}$ has been calculated in \cite{SMI1}
by fitting the width of the $\Lambda^0(1405)$ hyperon, treated as a
bound $K^-p$ state ${^1_{\bar{K}}}{\rm H}$.

We would like to make clear the title of our model namely ``The
phenomenological quantum field theoretic model''.  On the one hand our
model is ``phenomenological'', since we construct the wave function of
the KNC, the parameters of which are fixed in terms of the
experimental data on the width and mass of the $\Lambda(1405)$
hyperon. On the other hand our model is ``quantum field theoretic'',
since to the calculation of scattering amplitudes of the constituents
of the KNC we apply the chiral Lagrangian with chiral $SU(3)\times
SU(3)$ symmetry and derivative meson--baryon couplings, ChPT and the
large $N_C$ expansion.

The binding energy $\epsilon_{S^0} = -\,197.0\,{\rm MeV}$ and the
width $\Gamma_{S^0} = 16\,{\rm MeV}$, calculated in our model, agree
well with the experimental data $\epsilon^{\exp}_{S^0} =
-\,194.0^{+\,1.5}_{-\,4.4}\,{\rm MeV}$ and $\Gamma^{\exp}_{S^0} <
21\,{\rm MeV}$.  We have found that the decay mode $S^0 \to \Sigma^-
pn$ is dominant, whereas the decay modes $S^0 \to \Sigma^0 nn$ and
$S^0 \to \Lambda^0 nn$ are suppressed.  The calculation of the
amplitudes of elastic and inelastic $K^-pnn$ scattering we have
carried out using the chiral Lagrangian with the non--linear
realization of chiral $SU(3)\times SU(3)$ symmetry and derivative
meson--baryon couplings (see Appendix B of Ref.\cite{SMI1}). We have
kept the contributions of leading order in the large $N_C$ and chiral
expansion \cite{SMI1}. In more detail the experimental analysis of the
theoretical amplitude of the $S^0 \to \Sigma^- p n$ decay can be
obtained through the measurements of the invariant--mass spectrum
\cite{PK06}.  The theoretical invariant--mass spectrum of the $S^0 \to
\Sigma^- p n$ decay is given by Eq.(\ref{label3.4}) and represented in
Fig.\,3.

We would like to accentuate that our theoretical predictions for the
binding energy $\epsilon_{S^0} = -\,197.0\,{\rm MeV}$ and the width
$\Gamma_{S^0} = 16\,{\rm MeV}$ of the KNC $S^0(3115)$ agree well with
results, obtained by Akaishi and Yamazaki within the potential model
approach \cite{TDB2} (see also \cite{EDB2}): $\epsilon_{S^0} =
-\,191.0\,{\rm MeV}$ and $\Gamma(S^0 \to \Sigma^-pn) = 13\,{\rm MeV}$.

However one can show that in our model the nuclear matter density of
the KNC $S^0(3115)$ is smaller by factor 3 than $n_{S^0}(0) =
1.56\,{\rm fm}^{-3}$, predicted in the potential model approach
\cite{EDB2}.  Indeed, using the wave function of the KNC $S^0(3115)$
one can calculate the nuclear matter density at the center of mass of
the $K^-pnn$ system.  This gives
\begin{eqnarray}\label{label4.1}
 n_{S^0}(0) = 3\,\Big(\frac{\mu^3_1\omega^2_1\Omega_1}{\pi^3}\Big)^{1/2} = 
0.53\,{\rm fm}^{-3}.
\end{eqnarray}
The obtained result is larger by a factor 4 than the normal nuclear
matter density $n_0 = 0.14\,{\rm fm}^{-3}$, but it is smaller by
factor 3 than $n_{S^0}(0) = 1.56\,{\rm fm}^{-3}$, predicted by Akaishi
and Yamazaki.

Our approach for the analysis of the Kaonic Nuclear Clusters $K^-p$
(or $\Lambda^0(1405))$, $K^-pp$ and $K^-pnn$ can be applied to the
description of the Kaonic Nuclear Clusters $K^-ppn$ with isospin $I =
0$, $K^-ppn$ and $K^-ppp$ with isospin $I = 1$, $K^-NNNN$ and so on
\cite{EDB2}.

\section*{Appendix A: Wave function $\Phi_{S^0}$ of the KNC $S^0(3115)$}
\renewcommand{\theequation}{A-\arabic{equation}}
\setcounter{equation}{0}

In the momentum representation the wave function $\Phi_{S^0}$ of the
KNC $S^0(3115)$ is defined by \cite{SMI1}
\begin{eqnarray}\label{labelA.1}
  \Phi_{S^0}(\vec{q},\vec{k},\vec{Q}\,)_{J M} = \Phi_K(\vec{q}\,)\Phi_{p(nn)}(\vec{k}\,)
\Phi_{nn}(Q)\Psi_{JM}(\vec{Q}\,).
\end{eqnarray}
This wave function describes the $K^-pnn$ system with total momentum
$J = \frac{3}{2}$ and positive parity, where $\Phi_K(\vec{q}\,)$,
$\Phi_{p(nn)}(\vec{k}\,)$ and $\Phi_{nn}(\vec{Q}\,)$ are the wave
functions of harmonic oscillators \cite{SMI1}
\begin{eqnarray}\label{labelA.2}
 \Phi_K(\vec{q}\,) &=& \Big(\frac{64\pi^3}{\mu^3_1
    \omega^2_1\Omega_1}\Big)^{1/4}\,\exp\Big(-\,\frac{\vec{q}^{\;2}_{\perp}}{2\mu_1
    \omega_1} - \frac{\vec{q}^{\;2}_{\parallel}}{2\mu_1
    \Omega_1}\Big),\nonumber\\
  \Phi_{p(nn)}(\vec{k}\,) &=& \Big(\frac{4\pi}{\mu_2
    \omega_2}\Big)^{3/4}\,\exp\Big(-\,\frac{\vec{k}^{\,2}}{2\mu_2
    \omega_2}\Big),\nonumber\\
  \Phi_{nn}(\vec{Q}\,) &=& \Phi_{nn}(Q)\,Y_{1 m}(\vartheta_{\vec{Q}},\varphi_{\vec{Q}}) = \nonumber\\
 &=&\frac{8}{\sqrt{3}} \Big(\frac{\pi}{\mu_3
    \omega_3}\Big)^{5/4}\,|\vec{Q}|\,\exp\Big(-\,\frac{\vec{Q}^{\,2}}{2\mu_3
    \omega_3}\Big)\,Y_{1 m}(\vartheta_{\vec{Q}},\varphi_{\vec{Q}}),
\end{eqnarray}
where the $nn$ pair is in the P--wave state, $Y_{1
  m}(\vartheta_{\vec{Q}},\varphi_{\vec{Q}})$ is a spherical harmonic,
$m = 0,\pm1$ is a magnetic quantum number.  The frequencies and the
reduced masses are
\begin{eqnarray}\label{labelA.3}
\omega_1 &=& \sqrt{\frac{3}{2}\frac{\mu}{\mu_1}}\,\omega = 76\,{\rm MeV}\;,\;
\mu_1 = \frac{3 m_K m_N}{m_K + 3 m_N} = 420\,{\rm MeV}\nonumber\\
\Omega_1 &=& \sqrt{3}\,\omega_1 = 132\,{\rm MeV},\nonumber\\
\omega_2 &=&\sqrt{\frac{\mu}{\mu_2}}\,\omega = 51\,{ \rm MeV}\;,\;
\mu_2 = \frac{2}{3}\,m_N  = 630\,{\rm MeV},\nonumber\\
\omega_3 &=&  \frac{1}{2}\,\sqrt{\frac{\mu}{\mu_3}}\,\omega = 30\,{\rm MeV}\;,\;
\mu_3 = \frac{1}{2}\,m_N = 470\,{\rm MeV},
\end{eqnarray}
where $\omega = 64\,{\rm MeV}$ is the frequency of the longitudinal
oscillation of the $K^-$--meson relative to the $pp$ pair in the KNC
${^2_{\bar{K}}}{\rm H}$ \cite{SMI1} and $\mu = 2m_K m_N/(m_K + 2 m_N)
= 391\,{\rm MeV}$ is the reduced mass of the $K^-(pp)$ system,
calculated for $m_K = 494\,{\rm MeV}$ and $m_N = 940\,{\rm MeV}$. The
value $\omega = 64\,{\rm MeV}$ is obtained from the fit of the width
of the $\Lambda^0(1405)$ hyperon, treated as a bound $K^-p$ state
\cite{SMI1}.

The wave functions $\Phi_K(\vec{q}\,)$, $\Phi_{p(nn)}(\vec{k}\,)$ and
$\Phi_{nn}(\vec{Q}\,)$ are normalised by
\begin{eqnarray}\label{labelA.4}
\int \frac{d^3q}{(2\pi)^3}|\Phi_K(\vec{q}\,)|^2 =\int
\frac{d^3k}{(2\pi)^3}|\Phi_{p(nn)}(\vec{k}\,)|^2 = \int
\frac{d^3Q}{(2\pi)^3}|\Phi_{nn}(\vec{Q}\,)|^2 = 1.
\end{eqnarray}
This gives
\begin{eqnarray}\label{labelA.5}
  \int \frac{d^3q}{(2\pi)^3}\Phi_K(\vec{q}\,) &=& 
  \Big(\frac{\mu^3_1\omega^2_1\Omega_1}{\pi^3}\Big)^{\!1/4}, \int
  \frac{d^3k}{(2\pi)^3} \Phi_{p(nn)}(\vec{k}\,) = 
  \Big(\frac{\mu_2\omega_2}{\pi}\Big)^{\!3/4},\nonumber\\
  \int
  \frac{d^3Q }{(2\pi)^3}\,Q\,\Phi_{nn}(Q)  &=&  \frac{12\pi}{\sqrt{6}}\,
  \Big(\frac{\mu_3\omega_3}{\pi}\Big)^{\!5/4}.
\end{eqnarray}
The wave functions $\Phi_K(\vec{q}\,)$, $\Phi_{p(nn)}(\vec{k}\,)$ and
$\Phi_{nn}(\vec{Q}\,)$ describe oscillations of the $K^-$--meson
relative to the $pnn$ system, the proton relative to the $nn$ pair and
the neutrons in the $nn$ pair, respectively.  The $K^-pnn$ system with
the wave function (\ref{labelA.1}) possesses positive parity.

Since the total angular momentum of the KNC $S^0(3115)$ is $J =
\frac{3}{2}$ \cite{TS04,TDB2,EDB2}, the wave functions $\Psi_{JM}(\vec{Q}\,)$ of
the angular momentum of the KNC $S^0(3115)$ are defined by
\begin{eqnarray}\label{labelA.6}
  \Psi_{\frac{3}{2},+\frac{3}{2}} &=&\frac{1}{2}\,Y_{1,+1}\,\Big[\chi^{(1)}_{\frac{1}{2},+ \frac{1}{2}}\,
  \chi^{(2)}_{\frac{1}{2},- \frac{1}{2}} 
  + 
  \chi^{(1)}_{\frac{1}{2},- \frac{1}{2}}\,\chi^{(2)}_{\frac{1}{2},+ \frac{1}{2}}\Big]\,\chi^{(p)}_{\frac{1}{2},+ \frac{1}{2}}
  - \frac{1}{\sqrt{2}}\,Y_{1,0}\,\chi^{(1)}_{\frac{1}{2},+ \frac{1}{2}}\,\chi^{(2)}_{\frac{1}{2},+ \frac{1}{2}}
  \,\chi^{(p)}_{\frac{1}{2},+ \frac{1}{2}}
  \nonumber\\
  \Psi_{\frac{3}{2},+\frac{1}{2}} &=&\frac{1}{\sqrt{12}}\,Y_{1,+1}\,\Big[\chi^{(1)}_{\frac{1}{2},+ \frac{1}{2}}\,
  \chi^{(2)}_{\frac{1}{2},- \frac{1}{2}} 
  + \chi^{(1)}_{\frac{1}{2},- \frac{1}{2}}\,\chi^{(2)}_{\frac{1}{2},+ \frac{1}{2}}\Big]\,\chi^{(p)}_{\frac{1}{2},- \frac{1}{2}}
  +\,\frac{1}{\sqrt{3}}\,Y_{1,+1}\,\chi^{(1)}_{\frac{1}{2},- \frac{1}{2}}\,\chi^{(2)}_{\frac{1}{2},- \frac{1}{2}}\,
  \chi^{(p)}_{\frac{1}{2},+ \frac{1}{2}}\nonumber\\
  &-&\frac{1}{\sqrt{6}}\,Y_{1,0}\,\chi^{(1)}_{\frac{1}{2},+ \frac{1}{2}}\,\chi^{(2)}_{\frac{1}{2},+ \frac{1}{2}}\,
  \chi^{(p)}_{\frac{1}{2},- \frac{1}{2}} -\,\frac{1}{\sqrt{3}}\,Y_{1,-1}\,
  \chi^{(1)}_{\frac{1}{2},+ \frac{1}{2}}\,\chi^{(2)}_{\frac{1}{2},+ \frac{1}{2}}\,
  \chi^{(p)}_{\frac{1}{2},+\frac{1}{2}}, \nonumber\\
  \Psi_{\frac{3}{2},-\frac{1}{2}} &=&\frac{1}{\sqrt{3}}\,Y_{1,+1}\,
  \chi^{(1)}_{\frac{1}{2},- \frac{1}{2}}\,\chi^{(2)}_{\frac{1}{2},- \frac{1}{2}}\,\chi^{(p)}_{\frac{1}{2},- \frac{1}{2}}
  +\,\frac{1}{\sqrt{6}}\,Y_{1,0}\,\chi^{(1)}_{\frac{1}{2},- \frac{1}{2}}\,\chi^{(2)}_{\frac{1}{2},- \frac{1}{2}}\,
  \chi^{(p)}_{\frac{1}{2},+ \frac{1}{2}}\nonumber\\
  &-&\frac{1}{\sqrt{3}}\,Y_{1,-1}\,\chi^{(1)}_{\frac{1}{2},+\frac{1}{2}}\,\chi^{(2)}_{\frac{1}{2},+ \frac{1}{2}}\,
  \chi^{(p)}_{\frac{1}{2},- \frac{1}{2}}-\,\frac{1}{\sqrt{12}}\,Y_{1,-1}\,\Big[\chi^{(1)}_{\frac{1}{2},+ \frac{1}{2}}\,
  \chi^{(2)}_{\frac{1}{2},- \frac{1}{2}} 
  + \chi^{(1)}_{\frac{1}{2},- \frac{1}{2}}\,\chi^{(2)}_{\frac{1}{2},+ \frac{1}{2}}\Big]\,\chi^{(p)}_{\frac{1}{2},+\frac{1}{2}},
  \nonumber\\
  \Psi_{\frac{3}{2},-\frac{3}{2}} &=&\frac{1}{\sqrt{2}}\,Y_{1,0}\,
  \chi^{(1)}_{\frac{1}{2},- \frac{1}{2}}\,\chi^{(2)}_{\frac{1}{2},- \frac{1}{2}}\,\chi^{(p)}_{\frac{1}{2},- \frac{1}{2}} - 
  \,\frac{1}{2}\,Y_{1,-1}\,
  + \chi^{(1)}_{\frac{1}{2},- \frac{1}{2}}\,\chi^{(2)}_{\frac{1}{2},+ \frac{1}{2}}\Big]\,\chi^{(p)}_{\frac{1}{2},+\frac{1}{2}},
\end{eqnarray}
where $\chi^{(a)}_{\frac{1}{2},\pm\frac{1}{2}}$ with $a =1,2,p$ are
the spinorial wave functions of the neutrons and the proton,
respectively.


\begin{thebibliography}{9}
\bibitem{TS04}
T. Suzuki {\it et al.} (the KEK Collaboration), Nucl. Phys. A {\bf 754}, 375 (2005);
Phys. Lett. B {\bf 597}, 263 (2004).
\bibitem{EDB1}
M. Agnello {\it et al.} (the FINUDA Collaboration),
Phys. Rev. Lett. {\bf 94}, 212303 (2005).
\bibitem{TDB1}
Y. Akaishi and T. Yamazaki,
Phys. Rev. C {\bf 65}, 044005 (2002);
T. Yamazaki and Y. Akaishi,
Phys. Lett. B {\bf 535}, 70 (2002).
\bibitem{TDB2}
Y. Akaishi and T. Yamazaki,
Nucl. Phys. A {\bf 684}, 409c (2001);
A. Dot${\acute{\rm e}}$, Y. Akaishi, and T. Yamazaki,
Mod. Phys. Lett. A {\bf 18}, 120 (2003);
T. Yamazaki, A. Dot${\acute{\rm e}}$, and Y. Akaishi,
Phys. Lett. B {\bf 587}, 167 (2004);
A. Dot${\acute{\rm e}}$ {\it et al.},
Phys. Lett. B {\bf 590}, 51 (2004);
Y. Akaishi, T. Yamazaki, and A. Dot${\acute{\rm e}}$,
Nucl. Phys. A {\bf 738}, 168 (2004);
A. Dot${\acute{\rm e}}$, Y. Akaishi, and T. Yamazaki,
Nucl. Phys. A {\bf 738}, 372 (2004).
\bibitem{EDB2}
A. Andronic, P. Braun--Munzinger, and K. Redlich,
Nucl. Phys. A {\bf 765}, 211 (2006) and references therein.
\bibitem{SMI1} 
A. N. Ivanov, P. Kienle, J. Marton, and E. Widmann,
  nucl--th/0512037; Invited talk at the Workshop on ``Exotic hadronic
  atoms, deeply bound kaonic nuclear states and anti--hydrogen:
  present results, future challenges'' at ECT$^*$ in Trento, 19--24
  June 2006, hep-ph/0610201; A plenary talk at the Workshop on ``QCD:
  Facts and Prospects'' in Oberw\"olz, Austria, 10--16 October 2006,
  http://physik.uni-graz.at/itp/oberw/
\bibitem{BWL68}
B. W. Lee,
Phys. Rev. {\bf 170}, 1359 (1968).
\bibitem{JG83}
J. Gasser, 
Nucl. Phys. Proc. Suppl. {\bf 86}, 257 (2000) and references therein.
H. Leutwyler, PiN Newslett. {\bf 15}, 1 (1999);
B. Borasoy and B. Holstein,
Eur. Phys. J. C {\bf 6}, 85 (1999);
Phys. Rev. D {\bf 59}, 094025 (1999);
Phys. Rev. D {\bf 60}, 054021 (1999);
Ulf-G. Mei\ss ner, 
PiN Newslett. {\bf 13}, 7 (1997);
H. Leutwyler,
 Ann. of Phys. {\bf 235}, 165 (1994);
G. Ecker, 
Prog. Part. Nucl. Phys. {\bf 36}, 71 (1996); 
Prog. Part. Nucl. Phys. {\bf 35}, 1 (1995); 
Nucl. Phys. Proc. Suppl. {\bf 16}, 581 (1990);
E. Jenkins and A. V. Manohar,
Phys. Lett. B {\bf 225}, 558 (1991);
A. Krause,
Helv. Phys. Acta {\bf 63}, 1 (1990);
J. Gasser, M. Sainio, and A. $\check{\rm S}$varc,
Nucl. Phys. B {\bf 307}, 779 (1988);
J. Gasser, Nucl. Phys. B {\bf 279}, 65 (1987);
 J. Gasser and  H. Leutwyler, 
Nucl. Phys. B {\bf 250}, 465 (1985); 
Ann. of  Phys. {\bf 158}, 142 (1984);
Phys. Lett. B {\bf 125}, 321 (1983).
\bibitem{HL00}
R. Kaiser and H. Leutwyler,
Eur. Phys. J. C {\bf 17}, 623 (2000).
\bibitem{EK01}
M. L. M. Lutz and E. E. Kolomeitsev,
Found. Phys. {\bf 31}, 1671 (2001).
\bibitem{EW79} 
E. Witten, Nucl. Phys. B {\bf 160}, 57 (1979).
\bibitem{TE05}
T. E. O. Ericson and A. N. Ivanov,
Phys. Lett. B {\bf 634}, 39  (2006);
hep\,--\,ph/0503277.
\bibitem{AI99}
A. N. Ivanov,
 M. Nagy, and N. I. Troitskaya,
Phys. Rev.  {\bf C59}, 451 (1999).
\bibitem{PSI2}
H.--Ch. Schr\"oder {\it et al.},
Eur. Phys. J. C {\bf 21}, 473 (2001).
\bibitem{TE04}
T. E. O. Ericson, B. Loiseau, and  S. Wycech,
 Phys. Lett. B {\bf 594}, 76 (2004).
\bibitem{PK06}
P. Kienle, Y. Akaishi, and T. Yamazaki,
Phys. Lett. B {\bf 632}, 187 (2006).
\bibitem{IV4}
A. N. Ivanov {\it et al.}, Eur. Phys. J. A {\bf 23}, 79 (2005), 
nucl--th/0406053.
\bibitem{MN79}
M. M. Nagels {\it et al.},
Nucl. Phys. B {\bf 147}, 189 (1979);
O. Dumbrajs {\it et al.}, Nucl. Phys. B {\bf 216},
277 (1983).
\bibitem{FSI}
K. M. Watson,
Phys. Rev. {\bf 88}, 1163 (1952);
A. B. Migdal,
Sov. Phys. JETP {\bf 1}, 2 (1955);
D. S. Koltun and A. Reitan,
Phys. Rev. {\bf 141}, 1413 (1966),
H. Machner and J. Haidenbauer,
J. Phys. G: Nucl. Part. Phys. {\bf 25}, R231 (1999);
A. N. Ivanov {\it et al.}, nucl--th/0509055.
\bibitem{SAID}
R. A. Arndt, I. I. Strakovsky, and R. L. Workman,
Phys. Rev. C {\bf 62}, 034005 (2000), nucl--th/0004039;
http://gwdac.phys.gwu.edu
\end{thebibliography}
\end{document}